\documentclass{nature}

\usepackage[fleqn]{amsmath} 
\newcommand{\msun}{M_\odot}
\newcommand{\vect}{\mathbf}

\title{Gravitational detection of a low-mass dark satellite at cosmological distance}

\author{S. Vegetti$^1$, D. J. Lagattuta$^2$, J. P. McKean$^3$, M. W. Auger$^4$, C. D. Fassnacht$^{2}$,  L. V. E. Koopmans$^{5}$}
\begin{document}

\maketitle

\begin{affiliations}
\item Kavli Institute for Astrophysics and Space Research, Massachusetts Institute of Technology, Cambridge, MA 02139, USA
\item Department of Physics, University of California, Davis, CA 95616, USA
\item ASTRON, Oude Hoogeveensedijk 4, 7991 PD Dwingeloo, The Netherlands
\item Department of Physics, University of California, Santa Barbara, CA 93106, USA
\item Kapteyn Astronomical Institute, University of Groningen, P.O. Box 800, 9700 AV Groningen, The Netherlands
\end{affiliations}

\begin{abstract} 

The mass-function of dwarf satellite galaxies that are observed around Local Group galaxies substantially differs from simulations\cite{dummy1,dummy2,dummy3,dummy4,dummy5} based on cold dark matter: the simulations predict many more dwarf galaxies than are seen. The Local Group, however, may be anomalous in this regard\cite{dummy8,dummy9}. A massive dark satellite in an early-type lens galaxy at $z=0.222$ was recently found\cite{dummy10} using a new method based on gravitational lensing\cite{dummy6,dummy7}, suggesting that the mass fraction contained in substructure could be higher than is predicted from simulations. The lack of very low mass detections, however, prohibited any constraint on their mass function. Here we report the presence of a $\mathbf{1.9\pm0.1 \times 10^{8}}$~M$_{\odot}$ dark satellite in the Einstein-ring system JVAS~B1938+666 (ref. 11) at $z =0.881$, where M$_{\odot}$ denotes solar mass. This satellite galaxy has a mass similar to the Sagittarius\cite{dummy12} galaxy, which is a satellite of the Milky Way. We determine the logarithmic slope of the mass function for substructure beyond the local Universe to be $\mathbf{\alpha=1.1^{+0.6}_{-0.4}}$, with an average mass-fraction of $\mathbf{\langle f \rangle =3.3^{+3.6}_{-1.8}~\%}$, by combining data on both of these recently discovered galaxies. Our results are consistent with the predictions from cold dark matter simulations\cite{dummy13, dummy14,dummy15} at the 95 per cent confidence level,and therefore agree with the view that galaxies formed hierarchically in a Universe composed of cold dark matter.

\end{abstract}


The gravitational lens system JVAS~B1938+666 (ref. 11) has a bright infrared background galaxy at redshift 2.059 (ref. 16), which is gravitationally lensed into an almost complete Einstein ring of diameter $\sim 0.9$ arcseconds by a massive elliptical galaxy at redshift 0.881 (ref. 17). The bright, highly-magnified Einstein ring made this system an excellent candidate in which to to search for surface brightness anomalies caused by very low mass (dark matter) substructure in the halo around the high redshift {\sl elliptical} lens galaxy. The presence of a low-mass substructure (e.g.\ a luminous or dark satellite galaxy; also denoted as {\sl substructure} hereafter) in the lens galaxy can introduce a localized perturbation of the arc surface brightness distribution. Owing to the multiplicity of the gravitationally lensed images that form these arcs, these `surface brightness anomalies' can be analysed using a pixelized-lens modelling technique and used to gravitationally detect and quantify the total mass and position of the substructure, down to masses as low as $\sim$0.1\% of the mass of the lens inside the Einstein radius\cite{dummy6, dummy7}.
The lens system was imaged at 1.6 and 2.2 micron using the Near Infrared Camera (NIRC2) on the W.~M. Keck 10-m telescope in June 2010. The adaptive optics system was used to correct the incoming wavefront for the blurring induced by the atmosphere, providing a nearly diffraction-limited point spread function (PSF) with a full width at half maximum (FWHM) of $\sim$70 milli-arcseconds. Further details for the data sets and their image processing can be found in Supplementary Information.

A smooth parametric model for the lens potential was constrained by using the surface brightness emission from the Einstein ring of each data set independently, having first removed the smooth light contribution from the lensing galaxy. We represented the mass model using an ellipsoidal power-law, $\rho(r)\propto r^{-\gamma}$, where $\rho(r)$ is the combined luminous and dark matter density as a function of the ellipsoidal radius $r$. The best-fitting model was then fixed and further refined using local potential corrections defined on a regular grid, which are translated into surface density corrections with the Laplace operator. We found for both the 1.6 and 2.2 micron adaptive optics data sets that there was a significant positive density correction, which indicated the presence of a mass substructure (see Fig. 1 and Supplementary Information). Directly from the pixelized potential correction, we measured a substructure mass of $\sim 1.7 \times 10^{8} ~M_\odot$ inside a projected radius of 600 pc around the density peak.

\begin{figure}
\begin{center}
\setlength{\unitlength}{1cm}
\begin{picture}(6,9.5)
\put(-5,0){\includegraphics{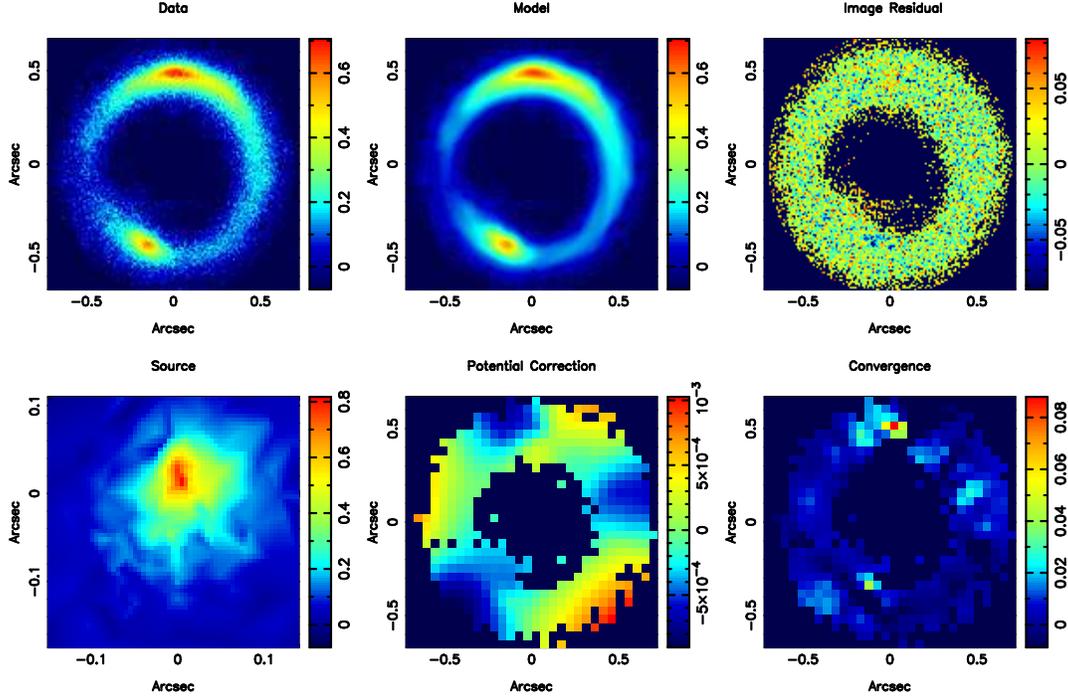}}
\end{picture}
\caption{The detection of a dark-matter dominated satellite in the gravitational lens system B1938+666 at redshift 0.881. The data shown here are at 2.2 micron and were taken with the W. M. Keck telescope in June 2010. Additional data sets at 1.6 micron, from the Keck telescope and the Hubble Space Telescope, are presented in the Supplementary Information. Top-left panel: the original data set with the lensing galaxy subtracted. Top-middle panel: the final reconstruction. Top-right panel: the image residuals. Bottom-left panel: the source reconstruction. Bottom-middle panel: the potential correction from a smooth potential required by the model to fit the data. Bottom-right panel: the resulting dimensionless projected density corrections. The total lensing potential is defined as the sum of an analytic potential for the host galaxy plus the local pixelized potential corrections defined on a Cartesian grid. The potential corrections are a general correction to the analytical smooth potential and correct for the presence of substructure, for large-scale moments in the density profile of the galaxy and shear. When the Laplace operator is applied to the potential corrections and translated into surface density corrections, the terms related to the shear and mass sheets become zero and a constant, respectively. A strong positive density correction is found on the top part of the lensed arc. Note that these images are set on a arbitrary regular grid that has the origin shifted relative to the centre of the smooth lens model by $\Delta$x = 0.024 arcsec and $\Delta$y = 0.089 arcsec. When this shift is taken into account the position of the density correction is consistent with the position of the substructure found in the analytic re-construction (see Supplementary Information).}
\label{data}
\end{center}
\end{figure}

As an independent test, we repeated the analysis of the 2.2 micron data set, which had the highest-significance positive density correction, with different models of the PSF, different data reduction techniques, different rotations of the lensed images, different models for the lens galaxy surface brightness subtraction, and different resolutions for the reconstructed source. We also analysed an independent data set taken at 1.6 micron with the Near Infrared Camera and Multi-Object Spectrograph (NICMOS) onboard NASA's {\it Hubble Space Telescope}. In total, we tested fourteen different models, and three different data sets that all independently led to the detection of a positive density correction at the same spatial position, although with varying levels of significance (see Supplementary Information). Differential extinction across the gravitational arc could also produce a surface brightness anomaly. However, the colour of the arc was found to be consistent around and at the location of substructure, ruling out the possibly that dust affected our results. 

We used an analytic model to determine the mass and the statistical significance of the substructure in the context of a physical model\cite{dummy6, dummy7}. In this analytic approach, a truncated pseudo-Jaffe model was used to parameterize the substructure mass and position, giving three extra free parameters. 
To obtain a good fit to the observed surface brightness distribution of the lensed background source at 2.2 micron, we found that a substructure was required at a position consistent with the positive density correction detected above. Assuming the substructure to be situated in the plane of the lens galaxy, we objectively compared the smooth and the substructure parametric models in terms of the Bayesian evidence, which is a measure of the probability of the data given a specific model (mariginalized over all parameters). Computing the marginalized evidence involved integrating over the multidimensional parameter space within predefined priors. In particular, we assumed that the substructure was equally likely to be located at any point in the lens plane and to have a mass between $4.0\times10^6~M_\odot$ and $4.0\times10^9~M_\odot$, the mass range where comparison with simulations was possible\cite{dummy13, dummy14}. See Supplementary Information for further details on the Bayesian evidence. We found that our nominal model, with a substructure of mass $M_{\rm sub} = 1.9 \pm 0.1 \times 10^{8}~M_\odot$ located at $(0.036\pm0.005,0.576\pm0.007)$ arcseconds relative to the lensing galaxy (in sky coordinates defined positive towards the north and the west, respectively), was preferred by a factor $e^{65}$ over a smooth model. This would heuristically correspond to a $12$-$\sigma$ detection of the substructure, if the posterior probability distribution function were Gaussian. This agrees well with the substructure mass found by the pixelized potential correction method described above, given the systematic and statistical uncertainties (see Supplementary information).

We also considered a model containing two substructures. The corresponding Bayesian evidence was significantly lower than the Bayesian evidence of the smooth model and the single substructure model. Above, we have quoted the statistical uncertainties for the substructure mass and position, but the uncertainties will also be related to several sources of systematic error, which are discussed in Supplementary Information. In general, the uncertainty in the substructure mass (under the assumption that it is located in the lens plane) is entirely dominated by the inference in the tidal truncation radius, which requires the substructure to be de-projected, yielding a systematic uncertainty of 0.45 dex.

On the basis of the detection of the substructure, we calculated the joint posterior probability function for the projected mass fraction of the halo that is made up of substructure\cite{dummy18}, $f$, and the slope, $\alpha$, of the substructure mass function, (where $dN \propto dMM^{-\alpha}$, where N is the number density of subhaloes per comoving volume and M is the substructure mass; see Supplementary Information). For the JVAS B1938+666 lensing galaxy {($M_{lens}=2.46\times10^{10}~M_\odot$ within $R_{E}=3.39$ kpc)}, we found that $f = 3.9^{+3.6}_{-2.4}~\%$ at the 68 per cent confidence level, using a uniform prior probability distribution for $\alpha$, for a substructure mass range between $4.0\times10^6 M_\odot$ and $4.0\times10^9 M_\odot$. For a Gaussian prior probability distribution for $\alpha$ centred at $1.9\pm0.1$, which is the slope of the mass function predicted from simulations\cite{dummy13,dummy14,dummy15}, $f = 1.5^{+1.5}_{-0.9}~\%$ at the 68 per cent confidence level. The predicted fraction of substructure from simulations within the same mass range and projected distance from the host halo centre is $\sim0.1_{-0.1}^{+0.3}\%$ (ref. 19), which is marginally smaller than the lower limit implied by our detection of a substructure at the 68 per cent confidence level, independent of the prior set on the slope of the mass function. However, these simulations model the formation of a Milky Way halo at $z=0$ and simulations of elliptical galaxies out to $z\sim 1$ must be made before a more quantitative conclusion can be drawn. 

Whereas flux-ratio anomalies of multiply imaged quasars have previously been used to statistically measure the level of substructure in cosmologically distant lens galaxies\cite{dummy20, dummy21,dummy22,dummy23}, these analyses have degeneracies when the substructures are dark and it is difficult to localize and measure the masses of individual substructures. Hence, the shape of the substructure mass function cannot be constrained and so far it has remained unclear whether the results from quasar flux ratio anomalies are in agreement or disagreement with numerical simulations\cite{dummy19}. Our new low-mass detection also allows us to constrain for the first time the slope of the substructure mass function for galaxies other than our own, when combined with the detection of the $\sim$18-fold more massive (i.e.\ $3.5\times 10^9 M_\odot$) substructure in the elliptical lens galaxy SDSS~J0946+1006 {($M_{\rm lens}=2.45\times10^{10}~M_\odot$ within $R_{\rm E}=4.60$ kpc)} at redshift 0.222 that was previously reported\cite{dummy10}. Combining both detections resulted in a slope of ${\alpha=1.1^{+0.6}_{-0.4}}$ for the mass function and an average mass-fraction of ${\langle f \rangle =3.3^{+3.6}_{-1.8}~\%}$ for elliptical galaxies at the 68 per cent confidence level. These results suggest that the slope of the mass function for elliptical galaxies is similar to that observed for Milky Way satellites, but that elliptical galaxies have higher mass fractions. 

So far, detailed studies of the mass and luminosity properties of substructures have been limited to the Milky Way\cite{dummy12, dummy24,dummy25,dummy26}, and to some extent also to M31 (ref. 27) at a distance of $\sim$1\,Mpc from the Milky Way. About thirty luminous satellite galaxies detected within the Milky Way virial radius ($\sim 250$\,kpc) are considered possible cases of cold dark matter substructure, albeit at a much lower abundance than is predicted from simulations\cite{dummy15}. Twenty-three of the Milky Way satellites have been found to have masses of $\sim10^7$~M$_{\odot}$ within a 300-pc projected radius, from observations of the dynamics of their stars\cite{dummy28}. This method for determining masses is limited to the Local Group owing to the faintness of the satellite galaxies, and can have a systematic error if the satellite is not in dynamic equilibrium or if there is foreground contamination. The three-dimensional mass of the JVAS B1938+666 substructure within a radius of 300 pc is $M_{300}= 7.2\pm0.5\times10^7~M_\odot$ ($M_{300}= 3.4\times10^7~M_\odot$ for a singular isothermal sphere model), which is comparable to the masses of the Milky Way satellites\cite{dummy28}, but is hosted by an elliptical galaxy with a velocity dispersion of $\sigma_{\rm{SIS}}= 187$~km\,s$^{-1}$ at a co-moving distance of $\sim$3 Gpc, corresponding approximately to a time when the Universe was half the age it is today. A 3-$\sigma$ upper-limit of $L_V < 5.4 \times 10^7 L_{V,\odot}$ was found for the luminosity of the substructure in the rest-frame $V$-band within the tidal radius $r_t = 440\pm5$~pc. This is about a factor four brighter than Sagittarius and Fornax\cite{dummy12}. The velocity dispersion of the satellite, based on the Einstein radius found from the best fitting model, is $\sigma_v \approx 16$~km\,s$^{-1}$, which corresponds to a circular velocity of $V_{\rm circ} \approx 27$~km\,s$^{-1}$. Its three-dimensional mass within 600 pc is $M_{600}=1.13\pm0.06\times10^8~M_\odot$. The observed properties (mass and circular velocity) are comparable to the Sagittarius dwarf galaxy of the Milky Way\cite{dummy12}.   

The mass-to-light ratios of low-mass satellites around the Milky Way have also been found to disagree with the expectations from simulations. It has been predicted that satellites with the luminosity of Fornax and Sagittarius should have a velocity dispersion that is a factor of two higher than has been observed ($\sim$ 16 km\,s$^{-1}$), which represents another problem for the cold dark matter paradigm\cite{dummy8}. In the case of the JVAS B1938+666 substructure, the 3$\sigma$ upper limit to the luminosity is in the range of the Fornax and Sagittarius satellites, and the resulting mass-to light ratio has a lower limit of $> 3.5~M_{\odot} / L_{V, \odot}$. Although this result is consistent with the mass-to-light ratios predicted from simulations, the limit on the luminosity of the substructure will need to be lowered by an order of magnitude before any meaningful comparison can be made. This will only be possible with the next generation of large optical telescopes, which is planned to come into operation by the end of the decade. 

\newpage
\noindent{\large{\bf Supplementary Information}}
\section *{Observations and data reduction}

The data used for this work were taken as part of the Strong-lensing at High Angular Resolution Program (SHARP) and will be discussed in detail by Lagattuta et al. (in preparation); here we present only a concise summary. We observed the B1938+666 lens system (ref. 11) on 2010 June 29 and 30, using
the Near Infrared Camera 2 (NIRC2) with the adaptive optics (AO) system
mounted on the Keck II telescope.  To achieve the best possible image
resolution, we used the NIRC2 ``narrow'' camera, with a 100 square
arcsecond field-of-view and a 0.01 arcseconds pixel scale.  The tip-tilt AO
correction was achieved by simultaneously observing an $R = 15$ magnitude
star that was located 18 arcseconds away from the target, while the secondary AO
correction was achieved with a 10-Watt, $ \lambda = 589~{\rm nm}$ sodium
laser guide star.  We collected $82 \times 180$ second exposures of B1938+666
in the K$^{\prime}$ band (2.2 micron), and $22 \times 300$ second exposures in
the H band (1.6 micron).

The two data sets were reduced independently. We first flat-fielded the data using an empirical sky flat that was created by median
stacking all of the raw science frames.  In order not to bias the
median sky flat, we made sure that all of the pixels that contained astronomical flux were masked out
during this process.  After flat-fielding, we subtracted a median sky
background from each image and then spatially resampled the images to
remove the geometric distortion.  The image registration was achieved using a pixel-based
cross-correlation technique. We used the sub-pixel shifts measured
from the cross-correlation to {\it drizzle} each frame onto a common coordinate system.  Finally, we median stacked the
reduced individual frames for a second time, in order to remove any bad-pixel
artefacts and cosmic ray relics.  This median stacked image was our final output frame.

We also made a separate reduction using the Center for Adaptive optics Treasury Survey (CATS) pipeline to test any systematics in the data reduction process.

For our analysis, we also used imaging data of B1938+666 that were taken with the {\it Hubble Space Telescope}  ({\it HST}) at 1.6 microns (F160W) with the Near Infrared Camera and Multi-Object Spectrograph (NICMOS). These data were previously published by King et al. (ref. 11) and we refer to that paper for further details. The data were gathered from the {\it HST} archive and reduced using the {\it multidrizzle} package$^{29}$ to produce the final images.

\section*{Lens galaxy surface brightness subtraction}
In order to remove the smooth light contribution of the lensing galaxy from the imaging data of B1938+666, we subtracted a model of the surface brightness profile that was fitted to the data using GALFIT$^{30}$. The model for the lensing galaxy was made using Sersic profiles. To prevent over-subtraction of the Einstein ring itself, we masked all of the pixels that contained any high surface-brightness flux from the background galaxy when we optimised the model. 

However, the seeing-limited component of the PSF is broad compared to the separation between the lensing galaxy and the source images and this causes some over-subtraction due to the GALFIT model fitting away part of the low surface brightness flux of the source images. We therefore implement a second subtraction scheme$^{31}$ to try to explicitly model all of the source light along with the lens light. We fit two Sersic components to both the lens and the source and use a singular isothermal ellipsoid model for the lensing kernel. This model is optimised and the best-fit lensing galaxy surface brightness model is then subtracted from the data, leaving just the source flux.

\section*{Dust in the lens galaxy}

Differential dust extinction within the lens galaxy can change the surface brightness of the lensed images and produce a surface brightness anomaly. In the case of B1938+666 the lens galaxy is a massive elliptical, which tend not to have large amounts of dust or show significant reddening. To estimate what effect, if any, dust has on the surface brightness of the lensed images, the colour of the 1.6 and the 2.2 micron data sets were measured at the location of the substructure, and at those parts of the arc away from the substructure position. We found the colour difference between these locations to be consistent within 0.01-mag, ruling out the effect of dust at a level exceeding the noise. We also note that if the anomaly were due to dust extinction, we would expect it to be considerably stronger in the H-band than in the K-band (note that the H-band is 0.8 micron in the rest-frame where dust can still have an effect) and hence the inferred masses to be very different. Instead they agree very well between the two bands.

\section*{Bayesian Lens modelling}
We modelled the lens system using a recently developed grid based modelling technique$^{9,10}$. This technique is fully embedded within the Bayesian statistical formalism that allows the most probable {\it a-posteriori} lens potential and background source to be determined. The method allows the level of structure in the source to be objectively inferred and for different model families to be compared/ranked. The method is grid based in the sense that both the lensed 
images and the background source surface brightness distribution are reconstructed on pixelized grids. 
While the image grid is a regular Cartesian grid defined by the data resolution, the source is reconstructed 
using a Delaunay tessellation; the source grid automatically adapts with the lensing magnification.
The source grid is built from the lens plane grid using the lens equation and a variable number of lens plane pixels
that are set on a case by case basis. The optimal number of lens plane pixels can be inferred through the Bayesian evidence.

Initially, we considered a smooth lens galaxy parametrized
as a softened power-law elliptical model with an external shear of strength, $\Gamma$ and angle, $\Gamma_{\theta}$.
The projected surface mass density depends on the following parameters:
the lens strength $b$, the position angle (defined north to east) $\theta$, the projected flattening $q$,
the centroid of the lens mass distribution $x_0$, $y_0$, the core radius $r_c$, and the density slope $\gamma$,
and is defined (assuming for clarity here a position angle of 90 degrees) as,
\begin{equation}
\kappa(r) = \frac{b}{2}\left(\frac{3-\gamma}{q}\right)^{\frac{\gamma-1}{2}}\left((x-x_0)^2+\frac{(y-y_0)^2}{q^2}+r_c^2\right)^{\frac{1-\gamma}{2}}.
\label{eqn:kappa}
\end{equation} 
The smooth lens model and corresponding background source were determined by optimising all of the above parameters with the exception of the core radius, which was set to the small value of $r_c=0.0001$ (we note that the smooth model was only used as a starting point for the modelling and the choice of setting $r_c$ to a small value is not relevant) in order to maximize the following Bayesian posterior probability,
\begin{equation}
\log P \left( \mathbf{s}\,|\,\mathbf{\eta},\lambda_s\right)= - \left(\frac{1}{2}\chi^2+\lambda_s^2 ||\mathbf{H}_s\mathbf{s}||^2_2\right)\ + {\rm const.},
\label{equ:posterior_s}
\end{equation}
In Eq. (\ref{equ:posterior_s}), $\mathbf{s}$ is the source surface brightness distribution and $\lambda_s$ its regularization,
$\bf{\eta}$ is a vector containing all of the model parameters introduced above. The source regularisation $\lambda_s$
sets the level of smoothness of the source and its value was objectively set by maximizing Eq. (\ref{equ:posterior_s}).
Several forms of regularisation, $\mathbf{H}_s$, can be adopted, and in this paper we chose to use a gradient regularisation.
Note that because of the presence of noise and because of the extended emission of the background source, 
lens modelling in terms of the Bayesian penalty function instead of a simple $\chi^2$ is essential$^{10,32}$.  

In order to quantify systematic effects on the lens modelling, we considered
three different data sets; the 1.6 and 2.2 micron data sets from the Keck telescope and the 1.6 micron data set from the {\it HST}. 
The most probable {\it a-posteriori} models for these data sets are listed in Tables 1 and 2, respectively, as $\rm{M_H}$, $\rm{M_K}$ and $\rm{M_{HST}}$.
Moreover, we considered twelve different models for the 2.2 micron data set, of which $\rm{M_{1\div4}}$ used a different point spread function (PSF) model, $\rm{M_{5}}$ used a higher number of lens plane pixels for the source grid, $\rm{M_{6}}$ had a source regularisation that is adaptive with the 
signal-to-noise ratio of the images, $\rm{M_{7}}$ used the CATS data-reduction pipeline, $\rm{M_{8}}$ used a data-set rotated by forty-five degrees, $\rm{M_{SIS}}$ assumed a singular isothermal density profile for the substructure, and $\rm{M_{9}}$, $\rm{M_{10}}$ and $\rm{M_{11}}$ used three different procedures for the lens galaxy surface brightness subtraction. We also considered four different models for the 1.6 micron data set from Keck using four different lens surface brightness subtraction models, $\rm{M_{12}}$, $\rm{M_{13}}$ and $\rm{M_{14}}$.
All of these different data sets and models led to a consistent lens potential and source surface brightness distribution, and will be presented by Lagattuta et al. (in preparation).
Once the best smooth lens model was found, mass substructure in the lens galaxy was searched for by allowing for local
potential corrections. The lens potential was then defined as the sum of a power-law elliptical model and pixelized corrections defined on a regular Cartesian grid,
\begin{equation}
\psi\left(\mathbf{x},\mathbf{\eta}\right)= \psi_{\rm smooth}\left(\mathbf{x},\mathbf{\eta}\right)+\delta\psi\left(\mathbf{x}\right).
\end{equation}

\noindent The Bayesian posterior,
\begin{equation}
\log P \left( \mathbf{s}\,|\,\mathbf{\eta},\lambda_s,\lambda_{\delta\psi}\right)=-\frac{1}{2}\chi^2-\lambda_s^2 ||\mathbf{H}_s\mathbf{s}||^2_2
-\lambda_{\delta\psi}^2 ||\mathbf{H}_{\delta\psi}\mathbf{{\delta\psi}}||^2_2\, + {\rm const.},
\label{equ:posterior_pot}
\end{equation}
was optimised in an iterative fashion for the source surface brightness distribution and the potential corrections,
while the smooth potential component was kept fixed at the best fitted value determined from the smooth lens model analysis described above. 

Substructures appeared as positive density corrections. This allowed the location and the number of possible substructures in a specific lens potential to be quantified. 
Previous simulations have shown that we can detect more than one substructure in the same system when located at a position where they can affect the lensed images$^{10}$. All three independent data sets and considered models showed a positive potential correction located at the same position, see for example Fig. 1 of the Letter and Figs. 2, 3 and 4 of this Supplementary Information.  Each data set had a different resolution, signal-to-noise, and was observed at a different wavelength, this excluded the possibility that our detection was an artefact related to the dynamic range of the images, dust, cosmic rays, bad pixels or other systematic effects. The reason that the substructure position is correlated with the brighter regions of the Einstein ring, is related to the fact that brighter regions are more sensitive to surface brightness perturbations and are therefore the regions where we are more likely to detect substructure with this method, if they are indeed present. It is possible that this lens system contains other equally low-mass or lower-mass substructure that we are not able to detect because of their less favourable location. 

In some cases, we also find a second smaller density peak in the lower left part of the images. We investigated this second peak and find it not to be significant, for the following reasons. First, this peak lies near the edge of the reconstruction region where the potential corrections are generally affected and non-zero because of regularisation effects$^{33}$. Second, the marginalised Bayesian evidence (see following section), disfavours a model with either a single substructure located at the position of the second density peak, or a model with two substructures, one for each peak, over a model with only a single substructure at the position of the main peak. Finally, an analytical model (see below) with a substructure forced at the position of the second peak gives a best recovered substructure mass consistent with zero. 

As a final comment we point out that potential correction maps are a general correction to the analytical smooth potential and correct therefore not only for the presence 
of substructure, but also for large-scale moments in the density profile of the galaxy and shear. The underlying smooth analytical models are different in each data set and model, and therefore require different levels of correction in terms of shear and mass sheet. This can cause the potential correction maps to look different for each data set or model. However, when the Laplace operator is applied to the potential correction map to get the convergence correction map the terms related to the shear and the mass sheets become zero or an irrelevant constant (the latter is in general forced to zero due to the curvature regularization), respectively.

Once the number of significantly detected substructure was determined to be one, we quantified its mass using a model where both the host lens galaxy and the mass substructure had a parametrized density profile. This was done to stabilise the mass determination and place it in the context of a parametric model that was used previously. As before, the lens galaxy was parametrized with a power-law elliptical profile [see Eq. (\ref{eqn:kappa})], while the substructure was parametrized using a pseudo-Jaffe truncated profile defined by,
\begin{equation}
\kappa(r) = \frac{b_{\rm{sub}}}{2}\left[r^{-1}-(r^2+r_t^2)^{-1/2}\right],
\end{equation}
where $r_t$ is the substructure tidal radius and $b_{\rm{sub}}$ is the lens strength; both are related to the main galaxy lens strength $b$ and to $M_{\rm{sub}}$ by $r_t =\sqrt{b_{\rm{sub}}b}$ and $M_{\rm{sub}}=\pi r_t b_{\rm{sub}} \Sigma_c$. This assumed that the projected and the true position of the substructure relative to the host lens galaxy were the same. While this can be on average considered correct, the uncertainty on the true position of the substructure is a considerable source of systematic error on the substructure mass, as we discuss in more detail below. Combining the last two relations leaves the total mass and position of the substructure on the lens plane as free parameters for the substructure model. By maximizing the Bayesian posterior probability, Eq. (\ref{equ:posterior_s}), in terms of $\eta$, $\lambda_s$, the substructure mass and substructure position, we found the best \emph{substructure} models listed in Tables 1, 2 and 3. All three data sets and different models gave consistent values for the lens galaxy and substructure parameters. Moreover, in all cases the recovered substructure position was consistent with the grid-based density correction previously identified. We also considered a model where the substructure had a singular isothermal profile ($\rm{M_{SIS}}$, which is equivalent to a pseudo-Jaffe with $r_t\rightarrow\inf$). We found it to be located at (0.04, 0.59) arcseconds relative to the lens galaxy, consistent within the uncertainties of all other models, and with a lens strength $b=0.003$ arcsec, which corresponded to a SIS velocity dispersion of $\sigma_v=15.6~\rm{km~s^{-1}}$.

\section*{The Substructure Mass from the Potential Grid Correction}

The mass determination of the substructure from the analytic (pseudo-Jaffe) model is robust and within the context of a physical model. A nearly
model-independent check of this mass determination, is to measure it directly from the grid-based potential correction. 
To ensure a correct comparison of the derived masses -- since some
of the substructure surface density could be due to a correction to the global smooth model -- we use the grid-based potential
solution starting from the same smooth mass model as in the fully analytic model. A straightforward estimator
comes from summing over pixels in the convergence map, which one determines through the Poisson equation, given in 
dimensionless units by $\nabla^{2} \delta \psi = 2 \delta \kappa$. For $3\times 3$ pixels centred on the convergence
peak, which is similar to a $\sim$600 pc aperture, we find a mass of $1.6 \times 10^{8} \msun$. This is in excellent agreement
with the analytic model, when correcting for the fact that this measurements is a projected mass and given the uncertainties. 
Because the convergence (i.e.\ surface density) map is directly based on the potential 
correction through finite differencing, its errors are correlated and harder to estimate. To reduce
the effect of finite differencing, we apply Gauss' theorem directly to the potential correction grid, by integrating the in-product of 
$\nabla \delta \psi$ with the normal vector (pointed inward) of a circular curve centred on the convergence peak. Tests show 
that this method suffers little from discretisation of the grid. This way, we find a mass of 
$1.7 \times 10^{8} \msun$ inside an exact 600\,pc projected radius, again in good agreement with 
the direct method and the analytic model. Second order differences could still be due to the choice of the substructure
density profile, but the agreement shows these effects to be small. We are therefore confident that both the grid-based {\sl and} 
analytic models give robust results, in agreement with each other within the errors. Throughout our analyses, for consistency, 
we will use only the mass determinations from the analytic model.

\section*{Model comparison}
Models with a different number of degrees of freedom can be consistently 
and objectively compared in the context of Bayesian statistics in terms of the marginalized Bayesian evidence. For a given model $\mathbf{M}$ and a regularisation form $\mathbf R$, the marginalized Bayesian evidence is defined as,
\begin{equation}
{\cal{E}}= P(\mathbf{M},\mathbf R\,|\,\mathbf d) \propto P(\mathbf d\,|\,\mathbf{M},\mathbf R)P(\mathbf{M},\mathbf R)\,.
  \end{equation}
This marginalized Bayesian evidence provides a measure of the probability of a specific model given the data $\mathbf d$.
If the prior probability $P(\mathbf{M},\mathbf R)$ is flat then different models can be compared according to their value of $P(\mathbf
  d\,|\,\mathbf{M},\mathbf R)$, which is related to the Bayesian evidence by,
   \begin{equation}
    P(\mathbf d\,|\,\mathbf{M},\mathbf R)=\int{d\lambda_s\, d\mathbf
      \eta \,P(\mathbf d\,|\,\mathbf \lambda_s,\mathbf \eta,\mathbf{M},\mathbf
      R) P(\lambda_s, \eta)}\,.
    \label{equ:evidence_integral}
  \end{equation}
$P(\mathbf\lambda_s,\mathbf \eta)$ is the prior probability distribution on all of the model parameters. 
In this Letter, we set all of the lens parameters in $\mathbf{\eta}$ to have uniform priors
centred around the best fitted values of the smooth $\rm{M_K}$ data set and as wide as 1 up to 2 orders of magnitude larger
than the dispersion estimated from all of the different $\rm{M_i}$ data sets and models, depending on the lens parameter.
We note that the priors for all models are identical.
The source regularisation constant had a uniform prior in the logarithmic space. 
The prior on the substructure position was uniform within the lens plane, 
that is, the substructure had an equal probability of being at any position in the lens plane and was not forced at any time to be
at the same position as the previously detected density correction. 
This can be considered as an extra test of the substructure detection.
The prior on the substructure mass was also uniform between $4\times10^{6}~M_{\odot}$ and $4\times10^{9}~M_{\odot}$.
To allow for a proper comparison, the priors had to be the same for both the smooth and substructured models.
A by-product of this integration is an exploration of the posterior probability, allowing for an error analysis of the model parameters and of the evidence itself. The results of this analysis for the 2.2 micron data set, which had the largest significance of all data sets, are given in Table 2, model $\rm M$. In Fig. 5, we show the posterior probability distribution for the substructure mass and position. We found that a model with a substructure at $(0.036\pm0.005,0.576\pm0.007)$ arcseconds relative to the lensing galaxy and with a mass of $\left(1.9\pm0.1\right)\times10^8 M_\odot$ was statistically preferred over a smooth model with a difference in the marginalized Bayesian evidence (Bayes factor) of $\Delta\log{\cal{E}}=\log{\cal{E}}_{smooth}-\log{\cal{E}}_{sub}=-65.10$, which corresponds to a $12~\sigma$ detection, if the posterior probability distribution function were Gaussian.
 
\section*{Systematic error on the substructure mass} 

Two major classes of systematic errors affect our measurement of the substructure mass. The first one is related to the modelling of the data and the data themselves, namely to the PSF modelling, the lens galaxy surface brightness subtraction, the data noise and resolution. The second is related to the de-projection of the substructure position.
As discussed above, in order to quantify the effect related to the first class of systematic error we have analysed three independent data sets and fourteen different models and found an uncertainty of $0.2\times 10^8 M_\odot$.

Two major assumptions about the physical position of the substructure were made when measuring its mass.
First, we assumed that the substructure is physically associated with the lens galaxy and therefore lies at the same redshift.
However, current high resolution simulations mostly concentrate on the formation of Milky-Way type of haloes and a precise quantification of the line-of-sight contamination on surface brightness anomalies for massive elliptical galaxies is still lacking.
Second, we assumed that the projected separation between the lens and the substructure is equivalent to the true separation.   
In general the true separation will be larger than the projected separation and the above inferred value represents a lower limit on the total mass, but is still a good representation of the mass within a fixed radius of $r=r_t$.

The posterior probability for $M_{\rm{sub}}$ is computed using the relation $M_{\rm{sub}}$ = $\pi r_t b_{\rm{ sub}}\Sigma_{c}$. If we use the tidal radius as the truncation radius then $r_t = r_s\left(\frac{M_{\rm{sub}}}{3 M}\right)^{1/3}$ (ref. 34) where $r_s$ is the true separation between the galaxy and the substructure, and $M$ is the mass of the galaxy within $r_s$. For B1938+666 this implies $M_{\rm{sub}}/M \approx \pi r_t b_{\rm{sub}}b^{-2}$ and hence $r_t = \frac{r_s^{3/2}}{b^{1/2}} \sqrt{\pi b_{sub}/3b}$.  If we assume that galaxy-scale substructures follow the total mass distribution, as is found in simulations and observations of more massive substructures$^{35,36}$, then $r_s$ will have a distribution that falls off approximately as $r^{-2}$. However, we also have the constraint that the projected separation is $r_p$, which yields a distribution of $r_s$ that goes as $\left(r \sqrt{(r/r_p)^2 -1 }\right)^{-1}$ (ref. 37). As $M_{sub} \propto r_s$, this implies the same functional form for the posterior probability of $M_{sub}$, neglecting the very small uncertainty on $b_{\rm{sub}}$; this posterior inference on $M_{sub}$ is shown in Fig. 6. We find that the de-projection yields a systematic uncertainty on the total mass of 0.45 dex at the 68 per cent confidence level.

\section*{Substructure luminosity}
We estimated an upper limit on the luminosity of the substructure by
calculating the quadrature sum of the pixel noise in a 11 pixel
$\times$ 11 pixel square aperture. The aperture width was chosen to match
the tidal diameter of the substructure. Two separate estimates of the
pixel noise were used, and the results were consistent. We found that
the 3-$\sigma$ upper limit on the observed apparent magnitude was
$K^\prime > 28$. Using a K-correction of $-$1.34, we found that the
rest-frame V-band absolute magnitude 3-$\sigma$ limit was $M_V > -14.5$,
which corresponded to a luminosity limit of $L_V < 5.4 \times 10^7 L_{V,\odot}$.
This limit is a factor of four brighter than the luminosities of the Fornax and Sagittarius
dwarf galaxies in the Local Group$^{12}$, and therefore similar luminosity galaxies at $z\sim0.9$ would not be observable in our deep adaptive optics imaging.

\section*{Statistics of detections}
The measurement of a mass substructure
can be statistically formalised and used to set constraints on the substructure projected mass
fraction $f$ and the substructure mass function slope $\alpha$ in lens galaxies$^{18}$. 
As we are sensitive only to substructures that are either massive enough or close enough to change the surface brightness distribution of the lensed arc, 
$f$ is defined as the projected dark matter mass fraction within an annulus of 0.1 arcseconds around the Einstein radius of the lensing galaxy, ($3.38\pm0.76$) kpc, where our sensitivity is maximized.  
We defined the probability that the observed substructure belongs to a parent population characterised
by a certain fraction and a certain mass function slope as,
\begin{equation}
P\left( \alpha,f ~|~ n_s,\mathbf{m}, \mathbf{p} \right) =\frac{{\cal L}\left( n_s,\mathbf{m}~|~\alpha,f,\mathbf{p}\right)P\left( \alpha,f ~| ~\mathbf{p}\right)}{P\left( n_s,\mathbf{m}~|~\mathbf{p}\right)},
\end{equation}
where $\mathbf{p} = \{M_{min}, M_{max}, M_{low}, M_{high}\}$. 

The likelihood probability function depends on the minimum and maximum mass a substructure can have ($M_{min}=4.0\times10^6M_\odot$ and $M_{max}=4.0\times10^9M_\odot$) as set by current numerical simulations$^{13,14}$, on the highest mass we can detect ($M_{high}=M_{max}$), and on the lowest mass we can detect ($M_{low}=3.0\times10^7M_\odot = 3.0\times\sigma_{M_{sub}}$).  Note that higher mass satellite galaxies would probably be visible in the 2.2 micron data set as luminous substructure$^{38}$, and hence, would be included in the lens model as a secondary galaxy. $P\left( \alpha,f ~| ~\mathbf{p}\right)$ is the prior probability function for $\alpha$ and $f$. We assumed for the mass fraction a uniform prior between 0 and 100 percent. For the prior on the mass function we considered two cases. In the first scenario, we defined a Gaussian prior centred on 1.9 that had a standard deviation of 0.1 (ref. 15; Fig. 7; black curves). In the second scenario, we chose a uniform prior between $-$1.0 and 3.0 (Fig. 6; blue curves). From the marginalized probabilities, $P\left(f\,|\,n_s=1,m=M_{\rm{sub}},\mathbf{p} \right)$ and $P\left( \alpha ~|~n_s=1,m=M_{\rm{sub}},\mathbf{p} \right)$, we derived $f = 3.9^{+3.6}_{-2.4}~\%$ and $\alpha = 1.0^{+0.8}_{-0.7}$ at the 68 per cent confidence level (CL) for a flat prior on $\alpha$, and $f = 1.5^{+1.5}_{-0.9}~\%$ and $\alpha = 1.9^{+0.1}_{-0.1}$ at the 68 per cent CL for a Gaussian prior. 

The independent detection of substructure in different lens galaxies can be combined by taking the product of their likelihood
such that, 
\begin{equation}
{\cal L}\left( \{n_s,\vect{m}\}~|~\alpha,f,\vect{p} \right) = \prod_{k=1}^{n_l}{\cal L}\left( n_{s,k},\vect{m}_k~|~\alpha,f,\vect{p} \right)\,,
\label{equ:multiple_lens}
\end{equation}  
which is then used with Eq. (8) to determine the joint probability of $\alpha$ and $f$.



\begin{table*}
\begin{center}
\begin{tabular}{cccccccc}
\hline
Parameter&$\rm{M_K}$&$\rm{M_1}$&$\rm{M_2}$&$\rm{M_3}$&$\rm{M_4}$&$\rm{M_5}$&$\rm{M_6}$\\
\hline
$b$ (arcseconds)					&0.452	&0.450	&0.450	&0.450	&0.451	&0.448	&0.449\\
$\theta$ (degrees)					&-22.29	&-21.32	&-21.87	&-20.66	&-20.21	&-23.30	&-22.65\\
$q$								&0.866	&0.868	&0.867	&0.870	&0.869	&0.870	&0.856\\
$\gamma$						&2.042	&2.043	&2.045	&2.046	&2.045	&2.049	&2.049\\
$\Gamma$						&0.015	&0.021	&0.018	&0.020	&0.018	&0.016	&0.015\\
$\Gamma_{\theta}$ (degrees)			&107.9	&108.4	&109.0	&108.9	&109.1	&109.3	&104.8\\
$M_{sub} ~(10^{8}~M_{\odot})$		&2.1		&2.6		&1.9 	&2.1		&2.1		&2.1		&1.8\\
$x_{sub}$ (arcseconds)				&0.034	&0.039	&0.031	&0.033	&0.033	&0.035	&0.043\\
$y_{sub}$ (arcseconds)				&0.571	&0.584	&0.544	&0.574	&0.570	&0.578	&0.588\\
$\log{\lambda_s}$					&-3.000	&-2.990	&-3.250	&-3.203	&-3.731	&-3.853	&0.009\\
\hline
\\
\end{tabular}
     \caption{Best-fit parameters for our lens model. $b$ is the model lens strength, and
        parametrizes the Einstein ring radius (in arcseconds).
        $\theta$ is the position angle of the mass of the lensing
        galaxy, $q$ is its axis ratio, $\gamma$
      is the slope of the mass profile, as parametrized in
      Eq. (\ref{eqn:kappa}), $\Gamma$ and $\Gamma_{\theta}$ are,
      respectively, the magnitude and position angle of an external
      shear source, $\lambda_s$ measures the smoothness of the flux
      distribution of the lensed background galaxy, $M_{sub}$ is the mass of the substructure within the tidal radius and 
      $x_{sub}$, and $y_{sub}$ are its coordinates relative to the lensing galaxy.}
\label{tab:modelling_results2}
\end{center}
\end{table*}
 
\begin{table*}
\begin{center}
\begin{tabular}{cccccccc}
\hline
Parameter&$\rm{M_7}$&$\rm{M_{8}}$&$\rm{M_{9}}$&$\rm{M_{10}}$&$\rm{M_{11}}$&$\rm{M_{SIS}}$&$\rm{M}$\\
\hline
$b$ (arcseconds)					&0.452	&0.417	&0.447	&0.443	&0.446	&0.448	&0.443$\pm$0.007\\
$\theta$ (degrees)					&-21.81	&-74.05	&-22.76	&-22.00	&-23.31	&-21.41	&-24.16$\pm$0.841\\
$q$								&0.868	&0.853	&0.859	&0.857	&0.857	&0.853	&0.872$\pm$0.003\\
$\gamma$						&2.044	&2.107	&2.033	&2.035	&2.037	&2.048	&2.058$\pm$0.013\\
$\Gamma$						&0.015	&0.008	&0.017	&0.014	&0.015	&0.018	&0.012$\pm$0.001\\
$\Gamma_{\theta}$ (degrees)			&106.8	&73.27	&97.91	&101.5	&99.92	&103.40	&113.2$\pm$2.521\\
$M_{sub}~(10^{8}~M_{\odot})$		&2.2		&1.9		&2.1		&2.1		&2.0		&0.026$^a$&1.9$\pm$0.1\\
$x_{sub}$ (arcseconds)				&0.036	&0.340	&0.030	&0.047	&0.053	&0.037	&0.036$\pm$0.005\\
$y_{sub}$ (arcseconds)				&0.578	&0.320	&0.553	&0.517	&0.507	&0.590	&0.576$\pm$0.007\\
$\log{\lambda_s}$					&-3.136	&-3.804	&-5.437	&-6.425	&-5.421	&-3.171	&-3.012$\pm$0.108\\
\hline
\\
\end{tabular}
\caption{Same as Table 1 for the $\rm{M_{7,8,9,10,11}}$ and $\rm{M_{SIS}}$ models. The mean values of the posterior probability distribution with relative standard deviation for each model parameter are given in the last column. $^a$ Note that the mass is defined within the Einstein radius of the substructure (0.003 arcseconds) and not the tidal radius.}
\label{tab:modelling_results2}
\end{center}
\end{table*}

\begin{table*}
\begin{center}
\begin{tabular}{cccccc}
\hline
Parameter&$\rm{M_H}$&$\rm{M_{12}}$&$\rm{M_{13}}$&$\rm{M_{14}}$&$\rm{M_{HST}}$\\
\hline
$b$ (arcseconds)				&0.423	&0.401	&0.407	&0.410	&0.438\\
$\theta$ (degrees)				&-24.19	&-24.76	&-26.40	&-27.22	&-23.06\\
$q$							&0.856	&0.858	&0.860	&0.859	&0.855\\
$\gamma$					&2.089	&2.095	&2.090	&2.084	&2.066\\
$\Gamma$					&0.014	&0.015	&0.016	&0.015	&0.016\\
$\Gamma_{\theta}$ (degrees)		&111.4	&110.5	&110.2	&114.1	&105.9\\
$M_{sub}~(10^{8}~M_{\odot})$	&2.0		&2.1		&2.0		&2.0		&1.9\\
$x_{sub}$ (arcseconds)			&0.030	&0.033	&0.059	&0.052	&0.038\\
$y_{sub}$ (arcseconds)			&0.555	&0.507	&0.529	&0.518	&0.522\\
$\log{\lambda_s}$				&-0.468	&-3.683	&-2.627	&-3.003	&1.642\\
\hline
\\
\end{tabular}
\caption{Same as Table 1 for the $\rm{M_{12,13,14}}$ models and the $\rm{M_{H,HST}}$ data sets.}

\label{tab:modelling_results2}
\end{center}
\end{table*}

\begin{figure}
\begin{center}
\setlength{\unitlength}{1cm}
\begin{picture}(6,10.2)
\put(-5,0){\includegraphics{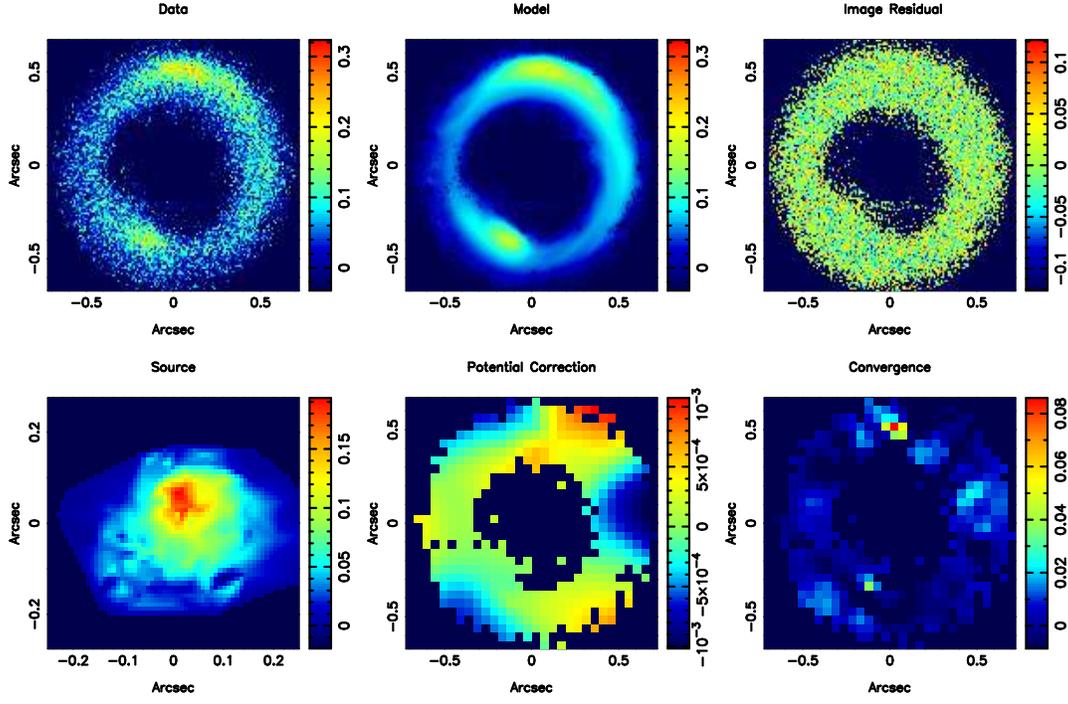}}
\end{picture}
\caption{The detection of a dark matter substructure as a positive pixelized density correction in the 1.6 micron Keck data set, $\rm{M_H}$. The top-right panel shows the original lens data, the middle one shows the final reconstruction while the top-left one shows the image residuals. On the second row the source reconstruction (left), the 
potential correction (middle) and the scale-free projected density corrections (right) are shown. The total lensing potential is defined as the sum of an analytic
potential for the host galaxy plus local pixelized potential corrections defined on a Cartesian grid. Using the Laplace operator, the potential corrections can be translated into density corrections {\bf ($\nabla^{2} \delta \psi = 2 \delta \kappa$)}. A strong positive correction is found on the top part of the lensed arc. Note that these images are set on a arbitrary regular grid that has the origin shifted relative to the centre of the smooth lens model by $\Delta$ x = 0.012 and $\Delta$ y = 0.091 arcseconds. When this shift is taken into account the position of the density correction is consistent with the position of the substructure in the analytic reconstruction.}
\label{highres}
\end{center}
\end{figure}

\begin{figure}
\begin{center}
\setlength{\unitlength}{1cm}
\begin{picture}(6,10.2)
\put(-5,0){\includegraphics{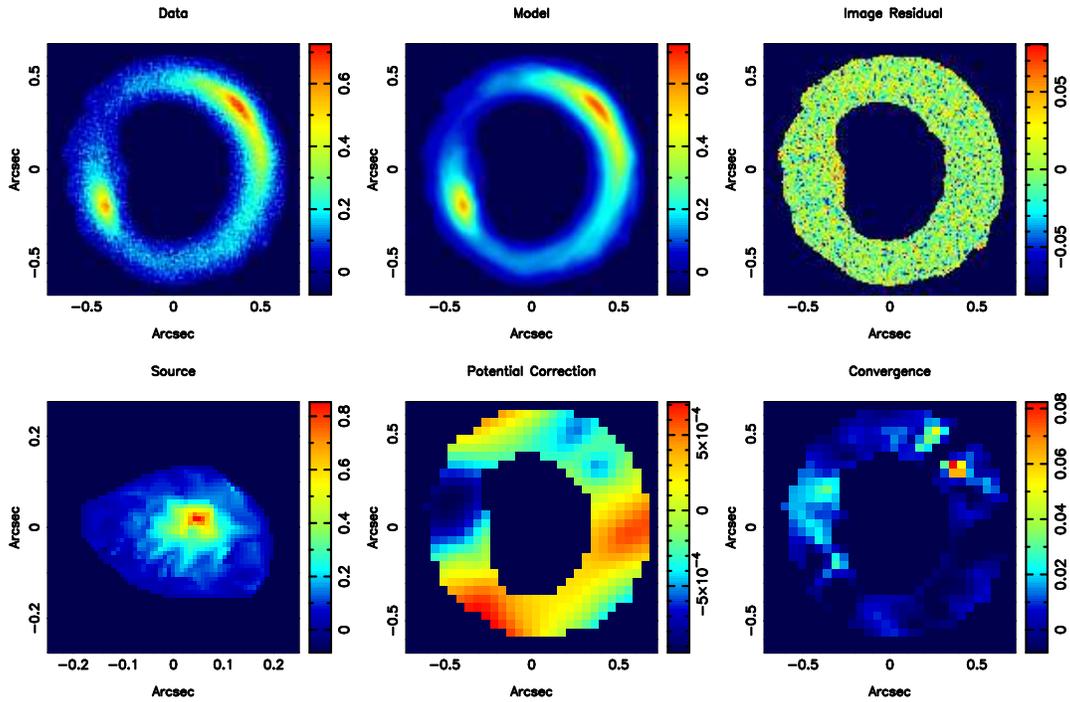}}
\end{picture}
\caption{The detection of a dark matter substructure as a positive pixelized density correction in the 2.2 micron Keck data set rotated by 45 degrees, $\rm{M_{8}}$. Note that these images are set on a arbitrary regular grid that has the origin shifted relative to the centre of the smooth lens model by $\Delta$ x = 0.062 and $\Delta$ y = 0.048 arcseconds along the x and y axis respectively. When this shift is taken into account the position of the density correction is consistent with the position of the substructure in the analytic reconstruction.}
\label{rotated}
\end{center}
\end{figure}

\begin{figure}
\begin{center}
\setlength{\unitlength}{1cm}
\begin{picture}(6,10.2)
\put(-5,0){\includegraphics{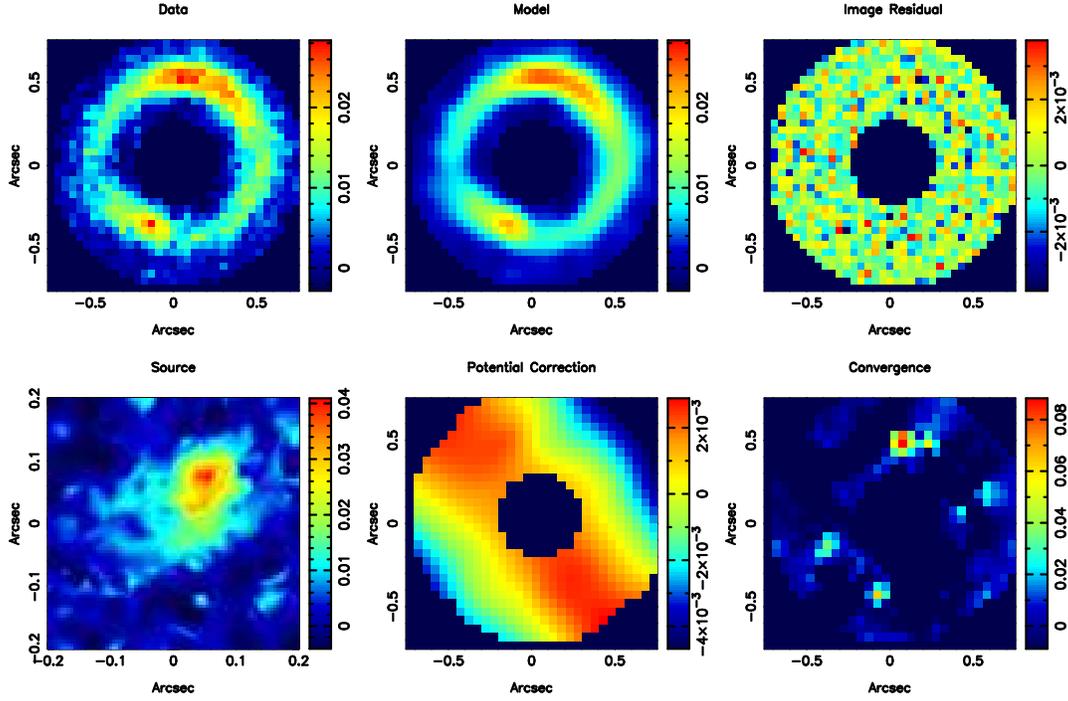}}
\end{picture}
\caption{The detection of a dark matter substructure as a positive pixelized density correction in the 1.6 micron {\it HST } data set, $\rm{M_{HST}}$. Note that these images are set on a arbitrary regular grid that has the origin shifted relative to the centre of the smooth lens model by $\Delta$ x = 0.001 and $\Delta$ y = 0.020 arcseconds along the x and y axis respectively. When this shift is taken into account the position of the density correction is consistent with the position of the substructure in the analytic reconstruction.}
\label{rotated}
\end{center}
\end{figure}

\begin{figure}
\begin{center}
\setlength{\unitlength}{1cm}
\begin{picture}(8,5.2)
\put(-3.5,0){\includegraphics{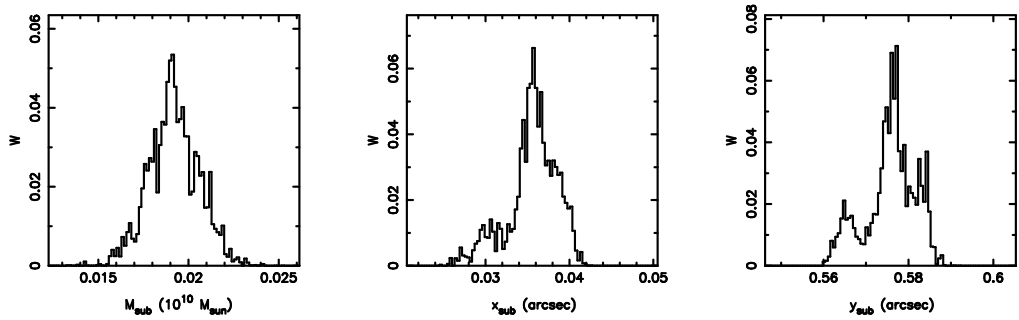}}
\end{picture}
\caption{Posterior probability distributions for the substructure mass (left) and position relative to the lens centre as obtained from
the Nested-Sampling evidence exploration for the 2.2 micron Keck data set.}
\label{data}
\end{center}
\end{figure}

\begin{figure}
\begin{center}
\setlength{\unitlength}{1cm}
\begin{picture}(6,5.2)
\put(-3,-3.5){\includegraphics{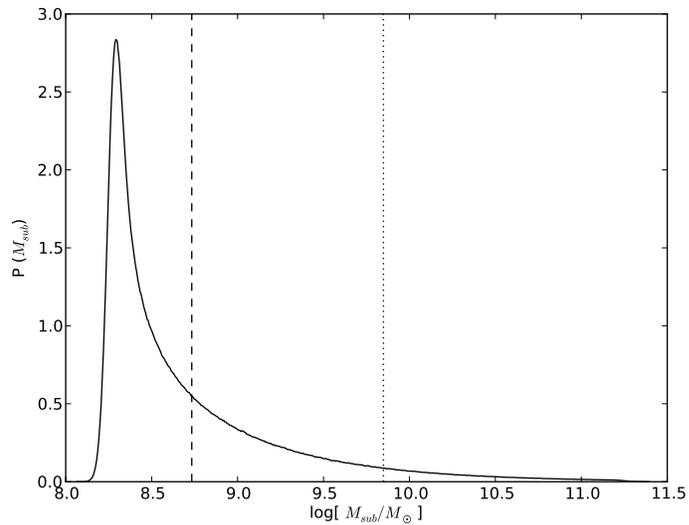}}
\end{picture}
\caption{Posterior on the inferred mass of the substructure, given the expected de-projection, with the intervals enclosing 68 per cent and 95 per cent of the posterior volume indicated by the vertical dashed and dotted lines, respectively.}
\label{data}
\end{center}
\end{figure}

\begin{figure}
\begin{center}
\setlength{\unitlength}{1cm}
\begin{picture}(6,7.5)
\put(-2.5,-0.5){\includegraphics{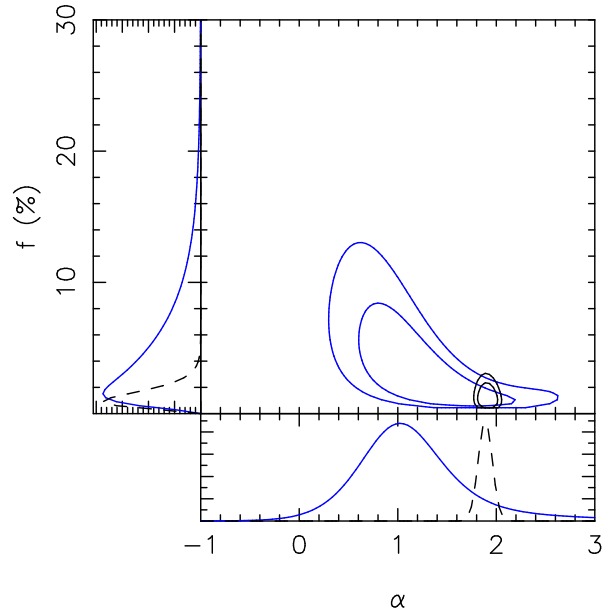}}
\end{picture}
\caption{The joint probability $P\left( \alpha,f ~|~ \{n_s,\vect{m}\},\vect{p} \right)$ contours and marginalized probabilities $P\left( f ~|~\{n_s,\vect{m}\},\vect{p} \right)$ and $P\left( \alpha ~|~\{n_s,\vect{m}\},\vect{p} \right)$ for a uniform prior (solid lines) and for  a Gaussian prior in $\alpha$ (dashed lines) as obtained by combining the detection presented in this paper and the detection in the lens system SDSS J0946+1006 (ref. 8).}
\label{prob}
\end{center}
\end{figure}

\newpage
\begin{addendum}

\item[Acknowledgements] Our results are based on observations made with the W. M. Keck Observatory and the {\it Hubble Space Telescope}. S.V. is supported by a Pappalardo Fellowship at the Massachusetts Institute of Technology, L.V.E.K. is supported (in part) through an NWO-VIDI program subsidy, D.J.L. and C.D.F. acknowledge support from the National Science Foundation. The authors are grateful to Phil Marshall for useful comments and feedback, and to the referees for a useful discussion that substantially improved the quality of the manuscript. 

\item[Author Contributions] S.V. and L.V.E.K. developed the gravitational imaging technique used for the detection of substructure. ~S.V. carried out the gravitational lens modelling of the data with help from L.V.E.K. and J.P.M.  ~S.V., L.V.E.K. and J.P.M. wrote the manuscript with comments from all of the authors. ~C.D.F. was the PI of the observing programme and was responsible, along with D.J.L., for acquiring the data. ~D.J.L. and M.W.A. reduced the data and performed the galaxy subtraction with help from C.D.F. ~M.W.A. calculated the systematic error on the substructure mass. ~C.D.F. calculated the galaxy luminosity.

\item[Author Information] Reprints and permissions information is available at www.nature.com/reprints. Correspondence and requests for materials should be addressed to S.V.~(email: svegetti@space.mit.edu)

\end{addendum}


\section*{References}
\begin{thebibliography}{1}

\bibitem{dummy1}
Kravtsov, A. Dark Matter Substructure and Dwarf Galactic Satellites. {\it Advances in Astronomy}, {\bf 2010}, 281913 (2010)

\bibitem{dummy2}
Kauffmann, G., White, S. D. M. \& Guiderdoni, B. The formation and evolution of galaxies within merging dark matter haloes. {\it Mon. Not. R. Astron. Soc.} {\bf 264}, 201-218 (1993)

\bibitem{dummy3}
Klypin, A., Kravtsov, A. V., Valenzuela, O. \& Prada, F. Where Are the Missing Galactic Satellites? {\it Astrophys. J} {\bf 522}, 82-92 (1999)

\bibitem{dummy4}
Moore, B., Ghigna, S., Governato, F., Lake, G., Quinn, T., Stadel, J. \& Tozzi, P. Dark Matter Substructure within Galactic Halos. {\it Astrophys. J} {\bf 524} L19-L22

\bibitem{dummy5}
Boylan-Kolchin, M., Bullock, J. S. \& Kaplinghat, M. Too big to fail? The puzzling darkness of massive Milky Way subhaloes. {\it Mon. Not. R. Astron. Soc.} {\bf 415}, L40-L44 (2011)

\bibitem{dummy8}
Boylan-Kolchin, M., Springel,V., White, S. D. M. \& Jenkins, A. There's no place like home? Statistics of Milky Way-mass dark matter haloes.  {\it Mon. Not. R. Astron. Soc.} {\bf 406}, 896-912 (2010)

\bibitem{dummy9}
Busha, M. T., Wechsler, R. H., Behroozi, P. S., Gerke, B. F., Klypin, A. A. \& Primack , J. R. Statistics of Satellite Galaxies Around Milky Way-Like Hosts. {\it Astrophys. J.}, submitted (2010)

\bibitem{dummy10}
Vegetti, S., Koopmans, L. V. E., Bolton, A., Treu, T. \& Gavazzi, R. Detection of a dark substructure through gravitational imaging.  {\it Mon. Not. R. Astron. Soc.} {\bf 408}, 1969-1981 (2010)

\bibitem{dummy6}
Koopmans, L. V. E. Gravitational imaging of cold dark matter substructures. {\it Mon. Not. R. Astron. Soc.} {\bf 363}, 1136-1144 (2005)

\bibitem{dummy7}
Vegetti, S., Koopmans, L. V. E. Bayesian strong gravitational-lens modelling on adaptive grids: objective detection of mass substructure in Galaxies. {\it Mon. Not. R. Astron. Soc.} {\bf 392}, 954-963 (2009)

\bibitem{dummy11}
King, L. J., Jackson, N., Blandford, R. D., Bremer, M. N., Browne, I. W. A., de Bruyn, A. G., Fassnacht, C., Koopmans, L. V. E., Marlow, D. \& Wilkinson, P. N. A complete infrared Einstein ring in the gravitational lens system B1938 + 666. {\it Mon. Not. R. Astron. Soc.} {\bf295}, L41-L44 (1998)

\bibitem{dummy12}
Strigari, L. E. et al. Redefining the missing satellites problem. {\it Astrophys. J.} 669, 676-683 (2007)

\bibitem{dummy13}
Diemand, J., Kuhlen, M. \& Madau, P. Dark matter substructure and gamma-ray annihilation in the Milky Way halo. {\it Astrophys. J.} {\bf  657}, 262-270 (2007)

\bibitem{dummy14}
Diemand, J., Kuhlen, M. \& Madau, P. Formation and evolution of galaxy dark matter halos and their substructure. {\it Astrophys. J} {\bf  667}, 859-877 (2007)

\bibitem{dummy15}
Springel, V., Wang, J., Vogelsberger, M., Ludlow, A., Jenkins, A., Helmi, A., Navarro, J. F., Frenk, C. S.\& White, S. D. M. The Aquarius Project: the subhaloes of galactic haloes. {\it Mon. Not. R. Astron. Soc.} {\bf 391} 1685-1711 (2008)

\bibitem{dummy16}
Riechers, D. A. Molecular Gas in Lensed $z >2$ Quasar Host Galaxies and the Star Formation Law for Galaxies with Luminous Active Galactic Nuclei. {\it Astrophys. J.} {\bf 730}, 108 (2011)

\bibitem{dummy17}
Tonry, J. L. \& Kochanek C. S. Redshifts of the Gravitational Lenses MG 1131+0456 and B1938+666. {\it Astrophys. J.} {\bf 119}, 1078-1082 (2000)

\bibitem{dummy18}
Vegetti, S., Koopmans, L. V. E. Statistics of mass substructure from strong gravitational lensing: quantifying the mass fraction and mass function.  {\it Mon. Not. R. Astron. Soc.} {\bf 400}, 1583-1592 (2009)

\bibitem{dummy19}
Xu, D. D., Mao, S., Wang, J., Springel, V., Gao, L., White, S. D. M., Frenk, C. S., Jenkins, A., Li, G. \& Navarro, J. F. Effects of dark matter substructures on gravitational lensing: results from the Aquarius simulations.  {\it Mon. Not R. Astron. Soc.} {\bf 398}, 1235-1253 (2009)

\bibitem{dummy20} Mao, S. \& Schneider, P. Evidence for substructure in lens galaxies?  {\it Mon. Not. R. Astron. Soc.} {\bf 295}, 587-594 (1998)

\bibitem{dummy21}
Dalal, N. and Kochanek, C. S. Direct Detection of Cold Dark Matter Substructure. {\it Astrophys. J.} {\bf 572}, 23-33 (2002)

\bibitem{dummy22}
Metcalf, R. B. \& Zhao, H. Flux Ratios as a Probe of Dark Substructures in Quadruple-Image Gravitational Lenses. {\it Astrophys. J.} {\bf  567}, L5-L8 (2002)
  
\bibitem{dummy23}
Keeton, C. R., Gaudi, B. S. \& Petters, A. O. Identifying Lenses with Small-Scale Structure. I. Cusp Lenses. {\it Astrophys. J.} {\bf  598}, 138-161 (2003)

\bibitem{dummy24}
Walker, M. G. et al. Velocity dispersion profiles of seven dwarf spheroidal galaxies. {\it Astrophys. J.} {\bf 667}, L53-L56 (2007)

\bibitem{dummy25}
Gilmore, G. et al. The observed properties of dark matter on small spatial scales. {\it Astrophys. J.} {\bf 663}, 948-959 (2007)

\bibitem{dummy26}
Simon, J. D. \& Geha, M. The kinematics of the ultra-faint Milky Way satellites: solving the missing satellite problem. {\it Astrophys. J.} {\bf 670}, 313-331 (2007)

\bibitem{dummy27}
Mateo, M. L. Dwarf galaxies of the Local Group. {\it Annu. Rev. Astron. Astrophys.} {\bf 36}, 435-506 (1998)

\bibitem{dummy28}
Strigari, L. E., Bullock, J. S., Kaplinghat, M., Simon, J. D., Geha, M., Willman, B. \& Walker, M. G. A common mass scale for satellite galaxies of the Milky Way. {\it Nature} {\bf 454} 1096-1097 (2008)
 
\bibitem{dummy29} 
Koekemoer, A.M., Fruchter, A.S., Hook, R.N. \& Hack, W. MultiDrizzle: An Integrated Pyraf Script for Registering, Cleaning and Combining Images.
{\it The 2002 HST Calibration Workshop : Hubble after the Installation of the ACS and the NICMOS Cooling System}, (2002)

\bibitem{dummy30}
Peng, C. Y., Ho, L. C., Impey, C. D. \& Rix, H. W. Detailed Structural Decomposition of Galaxy Images.
{\it Astronomical Journal}, {\bf 124}, 266-293 (2002)

\bibitem{dummy31}
Auger, M. W., Treu, T., Brewer, B.J. \& Marshall, P. J. A compact early-type galaxy at z= 0.6 under a magnifying lens: evidence for inside-out growth. {\it Mon. Not R. Astron. Soc.}, {\bf 411}, L6-L10 (2011)

\bibitem{dummy32}
Suyu, S. H., Marshall, P. J., Hobson, M. P. \& Blandford, R. D. A Bayesian analysis of regularized source inversions in gravitational lensing.  {\it Mon. Not R. Astron. Soc.}, {\bf 371}, 983-998 (2006)

\bibitem{dummy33}
Suyu, S. H., Marshall, P. J., Blandford, R. D., Fassnacht, C. D., Koopmans, L. V. E., McKean, J. P. \& Treu, T. Dissecting the Gravitational Lens B1608+656. I. Lens Potential Reconstruction. {\it Astronomical Journal}, {\bf 691}, 277-298 (2009)

\bibitem{dummy34}
Binney, J. \& Tremaine, S. Galactic Dynamics. {\it Princeton University Press} (2009)

\bibitem{dummy35}
Chen, J. The galaxy cross-correlation function as a probe of the spatial distribution of galactic satellites. {\it  Astronomy \& Astrophysics }, {\bf 494}, 867-877 (2009)

\bibitem{dummy36}
Nierenberg, A. M., Auger, M. W., Treu, T., Marshall, P. J. \& Fassnacht, C. D. Luminous Satellites of Early-type Galaxies. I. Spatial Distribution. {\it Astrophys. J.}, {\bf 731},  44-61 (2011)

\bibitem{dummy37}
Suyu, S. H. \& Halkola, A. The halos of satellite galaxies: the companion of the massive elliptical lens SL2S J08544-0121. {\it Astronomy \& Astrophysics}, {\bf 524}, A94 (2010)

\bibitem{dummy38}
More, A., McKean, J. P., More, S., Porcas, R. W., Koopmans, L. V. E. \& Garrett, M. A. The role of luminous substructure in the gravitational lens system MG 2016+112. {\it Mon. Not R. Astron. Soc.}, {\bf 394}, 174-190 (2009)

\end{thebibliography}
\end{document}